\documentclass[12pt]{iopart}
\usepackage{graphicx}
\usepackage{upgreek}
\usepackage{iopams}
\usepackage{textcomp}
\usepackage{setstack}

\begin{document}

\title{Implanted Bottom Gate for Epitaxial Graphene on Silicon Carbide}

\author{D Waldmann$^1$, J Jobst$^1$, F Fromm$^2$, F Speck$^2$, T Seyller$^2$, M Krieger$^1$ and H B Weber$^1$}

\address{$^1$Lehrstuhl f\"ur Angewandte Physik, Friedrich-Alexander Universit\"at Erlangen-N\"urnberg, Staudtstr. 7, 91058 Erlangen, Germany}
\address{$^2$Lehrstuhl f\"ur Technische Physik, Friedrich-Alexander Universit\"at Erlangen-N\"urnberg, Erwin-Rommel-Str. 1, 91058 Erlangen, Germany}

\ead{heiko.weber@physik.uni-erlangen.de}
\begin{abstract}
We present a technique to tune the charge density of epitaxial graphene via an electrostatic gate that is buried in the silicon carbide substrate. The result is a device in which graphene remains accessible for further manipulation or investigation. Via nitrogen or phosphor implantation into a silicon carbide wafer and subsequent graphene growth, devices can routinely be fabricated using standard semiconductor technology. We have optimized samples for room temperature as well as for cryogenic temperature operation. Depending on implantation dose and temperature we operate in two gating regimes. In the first, the gating mechanism is similar to a MOSFET, the second is based on a tuned space charge region of the silicon carbide semiconductor. We present a detailed model that describes the two gating regimes and the transition in between. 

\end{abstract}

\maketitle

\section{Introduction}
The fascination of graphene lies in many scientific and technological specialties, including a linear dispersion relation, a nontrivial structure of the electronic wave function and unprecedented charge carrier mobilities \cite{Nov05, Bol08a, Dea10}. Furthermore it is a well-defined two dimensional electron gas residing openly at the surface. Hence it can be manipulated chemically and physically with, in principle, atomic resolution. Graphene provides an opportunity to investigate electronic transport and surface science spectroscopy on the same material \cite{Ber04a, Sch07, Mar07, Bos10, Job10}. Varying the charge density in graphene permits the exploration of many phenomena \cite{Nov05, Mar07, Sta09}. For the commonly used exfoliated graphene, this can easily be achieved by employing the silicon substrate as an electrostatic bottom gate and its oxide as the dielectrics. However, exfoliated graphene flakes have limitations in size and reproducibility.

\textit{Epitaxial graphene} has turned out to share the interesting properties of exfoliated graphene and is routinely fabricated in excellent qualities on large areas \cite{Ber06, Emt09}. Since SiC is a wide-bandgap semiconductor material, the use of the substrate as bottom gate is manifest, however technically not easy to achieve. Both highly insulating material and well conducting gate electrode have to be monolithically implemented, and this setup has further to serve as source for graphene growth. As a work-around, in some surface science experiments the charge carrier density in epitaxial graphene is controlled by applying specific ad-atoms  as a static top gate \cite{Oht06,Che07}.

We present a setup for bottom gated epitaxial graphene by means of ion implantation into a semiinsulating SiC wafer with subsequent graphene growth. In this way, a conductive layer is created which is separated from the graphene by an insulating region. In \cite{Wal11} we showed that with this setup the charge carrier density in so called \lq quasi-freestanding monolayer graphene\rq\ (QFMLG) obtained via hydrogen intercalation \cite{Spe10, Spe11} can be efficiently tuned. By adjusting the implantation energy $E_I$ and the implantation dose $D_\mathrm{I}$, the capacitance of the bottom gate as well as the optimum working temperature can be tailored. In this article we want to complement the previous publication \cite{Wal11} by (i) investigating the practical limits of implantation doses and a comparison of the implantation species nitrogen and phosphor and  (ii) explaining the operation principle in detail by a specifically adapted model, which is (iii) supported by experiments under UV illumination.
A striking feature in reference \cite{Wal11} was the observation of two different operation regimes. In the so called implanted plate capacitor (IPC) regime, that is observed for low temperatures, the situation is equivalent to a simple plate capacitor and the capacitance does not depend on gate voltage $V_\mathrm{g}$. In the Schottky capacitor (SC) regime, that is observed for elevated temperatures, the physics is governed by the graphene/SiC Schottky contact. The SC regime shows improved gate coupling and a dependence of the capacitance on $V_\mathrm{g}$. The model in section \ref{modeling} describes the two gating regimes as well as the transition in between. UV illumination provides a control parameter to force the system into the SC regime, even at low temperature (see section \ref{controlling}).

\section{Sample preparation and experimental details}
\begin{figure}
    \centering
        \includegraphics[width=9cm]{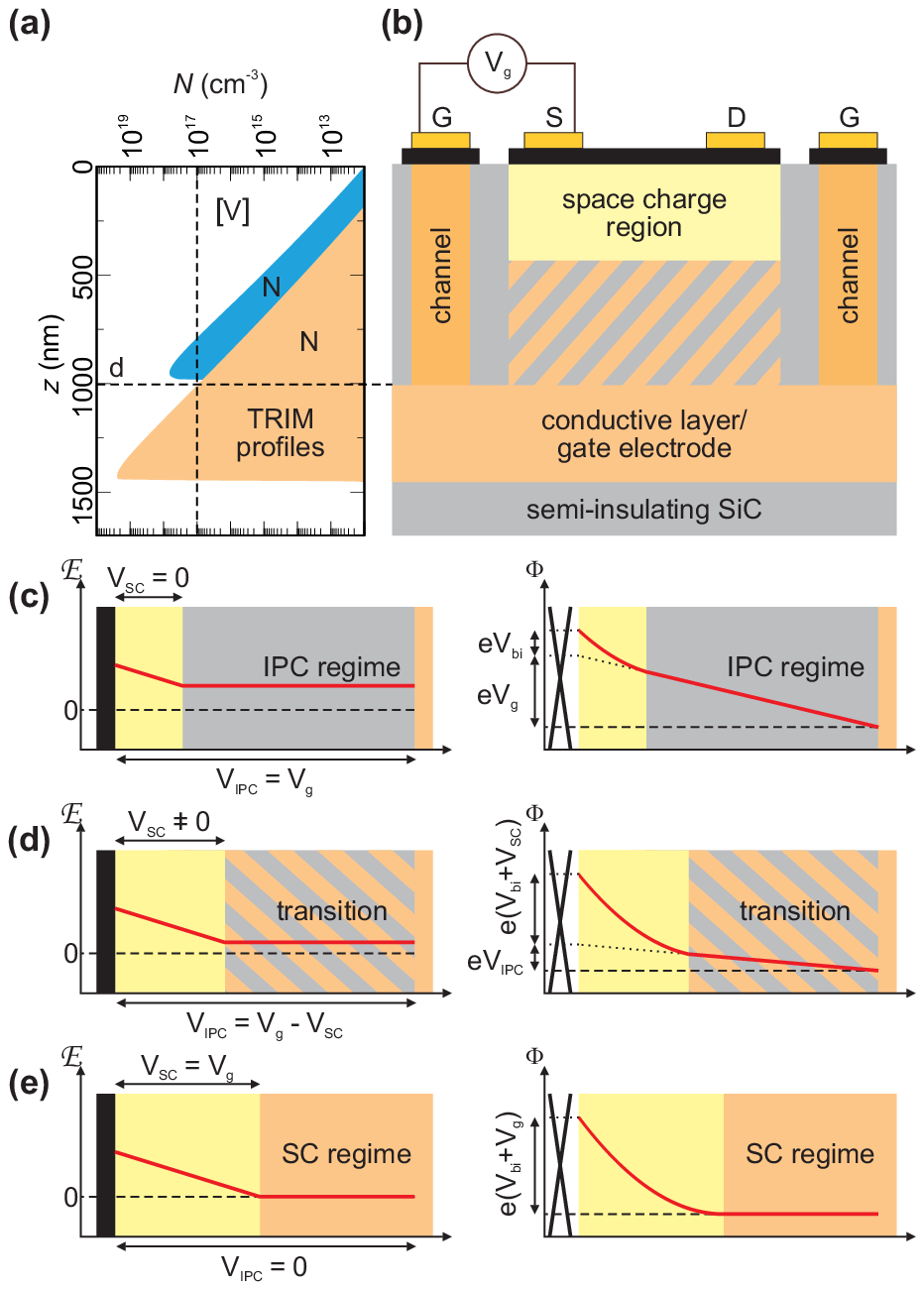}
    \caption{Scheme of the bottom gate. The conductive gate layer is created via implantation of nitrogen or phosphor ions in a depth $d$ below the SiC surface. (a) TRIM simulation of the implantation profiles of S1 (blue) and S3 and S4 (orange). The SiC is conducting where the implanted dose exceeds the vanadium compensation $[V]$ (vertical, dotted line) and insulating elsewhere. The horizontal, dotted line indicates $d$ for S3 and S4. (b) Setup of the bottom gate with source (S), drain (D) and gate (G) electrodes on graphene; conductive gate layer and implanted conductive channels. The area between the space charge region and the gate layer is insulating in the IPC regime whereas it is conducting in the SC regime. (c)--(e) Electric field and band diagram for this region in the different regimes. (c) In the IPC regime a constant electric field created by $V_\mathrm{g}$ is superimposed onto the built-in field of the depletion layer. (d) At the transition to the SC regime the constant electric field becomes smaller and the voltage drop along the space charge region increases. (e) In the SC regime the conductive layer extends up to the depletion layer. The whole gate voltage drops along the depletion layer.}
    \label{Schema}
\end{figure}

Our devices were fabricated on commercial on-axis, semi-insulating 6H-SiC (0001) wafers with a vanadium (V) compensation of $[V]\approx 10^{17}$\,cm$^{-3}$. The samples were hydrogen etched at $T=1520$\,\textcelsius\ to remove surface polishing damage similar to the process described in \cite{Emt09}. The conductive layer used as the bottom gate was created by means of ion implantation of nitrogen (N) or phosphor (P), which are shallow donors in SiC. The implantation profile that can be calculated using the Monte-Carlo simulation TRIM (Transport of Ions in Matter) \cite{Zie85} has its maximum at $z_{\mathrm{max}}=$1.0--1.4\,$\upmu$m depending on the implantation energy $E_\mathrm{I}$ and drops exponentially towards the surface. The SiC becomes conductive at a depth $d$ and below, where the doping concentration overcompensates the V concentration. This leads to the simultaneous formation of a conductive gate electrode with a maximum donor concentration $[D]_{\mathrm{gate}}$ and an \lq insulating layer\rq\ with thickness $d$ between the gate electrode and the surface (figure \ref{Schema}). Here, we report on five samples (S1 to S5) with different implantation doses, energies and species (see table 1).
\begin{table}
\caption{\label{tabone}Implantation species,  dose $D_\mathrm{I}$ and energy $E_\mathrm{I}$, the distance from the gate to the surface $d$, the theoretical and measured capacitance in the IPC regime and the typical breakdown voltage at room temperature $V_\mathrm{bd}$ for the presented samples.} 

\begin{indented}
\lineup
\item[]
\begin{tabular}{@{}lccccccc}
\br                             
&element&$D_\mathrm{I}$ & $E_\mathrm{I}$ & $d$ &$C_{\mathrm{IPC}}$ & $C_{\mathrm{IPC,m}}$&$V_\mathrm{bd}$\cr 
&& (cm$^{-2}$)&(MeV)&($\upmu$m)&(nF cm$^{-2})$&(nF cm$^{-2}$)&(V)\cr
\mr
S1&N&$8.0\cdot10^{12}$&$1.0$&0.8&11&11.4&$> 160$\cr
S2&N&$3.3\cdot10^{13}$&$1.0$&0.7&12&12.5&$\approx 100$\cr 
S3&N&$3.0\cdot10^{14}$&$2.0$&1.0&8.5&6.8&$\approx 20$\cr 
S4&N&$3.0\cdot10^{14}$&$2.0$&1.0&8.5&8.6&$\approx 20$\cr
S5&P&$8.0\cdot10^{14}$&$2.0$&0.6&14&8.5&$\approx 15 $\cr  
\br
\end{tabular}
\end{indented}
\end{table}
The bottom gate is connected to the surface with two vertical channels as described in \cite{Wal11}.
For the activation of the implanted dopants the samples were heated for 15--30\,min at temperatures between $T=1550$\,\textcelsius\ and $T=1700$\,\textcelsius\ followed by a 90\,s hydrogen etch at $T=1520$\,\textcelsius\ to remove surface damage caused by the implantation. For the growth of epitaxial graphene on SiC (0001) the process described in \cite{Spe10} is used to obtain hole doped QFMLG that is decoupled from the substrate by hydrogen intercalation. As shown in \cite{Wal11} only due to the intercalation step that saturates dangling bonds at the SiC/graphene interface \cite{Spe10}, the charge carrier density in the graphene can be tuned using the bottom gate.  Using standard e-beam lithography and reactive ion etching graphene was patterned in Hall bar structures similar to the process described in \cite{Job10}.

Micro-Raman spectra were measured under ambient conditions using a triple spectrometer with a liquid N$_2$ cooled CCD-detector. For excitation a frequency doubled Nd:YVO$_4$ laser with a wavelength of 532 nm was focused on the sample by a 100$\times$ objective.

Transfer characteristics and Hall measurements were either performed in a continuous-flow cryostat using magnetic fields of $\pm0.66$\,T at temperatures between  29\,K and 300\,K or in a Oxford cryostat with a minimal temperature of 1.4\,K and a magnetic field of up to 8\,T. The AC differential capacitance $C_{\mathrm{diff,AC}}$ measurements were performed using a LCR meter at a frequency $f=1$\,kHz (PSM1735 NumetriQ). The DC capacitance $C_{\mathrm{DC}}$ was obtained by measuring the Hall resistance $R_\mathrm{H}$ at gate voltages $V_\mathrm{g}=0$\,V and $V_\mathrm{g}=3$\,V with
\begin{equation}
C_{\mathrm{DC}}=\frac{1}{3\,\mathrm{V}}\left(\frac{1}{R_\mathrm{H}(V_\mathrm{g}=0\,{V})}-\frac{1}{R_\mathrm{H}(V_\mathrm{g}=3\,\mathrm{V})}\right)
\end{equation}
For the UV illumination a commercial UV-Diode with a wavelength of $\lambda=390$\,nm and an optical power of $P=2.2$\,mW was mounted roughly 1\,cm above the sample.

\section{Experimental results and discussion}
\begin{figure}
    \centering
        \includegraphics[width=9cm]{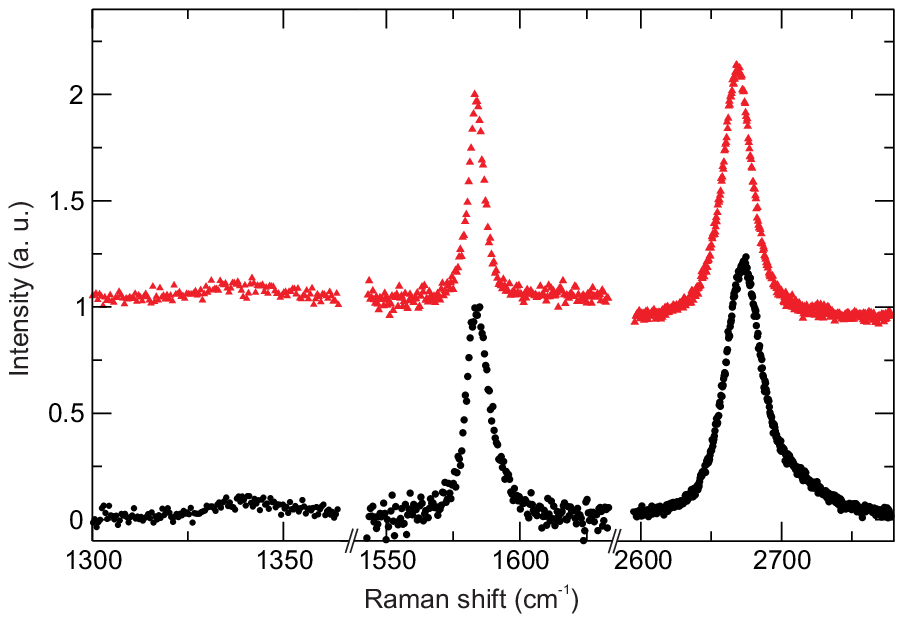}
    \caption{Raman spectra of QFMLG grown on pristine (circles) and implanted SiC (sample S4) (triangles). To remove the contribution of the substrate a SiC reference spectrum was subtracted. The spectra show the D-, G- and 2D band at 1340\,cm$^{-1}$, 1584\,cm$^{-1}$ and 2670\,cm$^{-1}$, respectively. The line widths of the G-peak are 6\,cm$^{-1}$ for S4 and 10\,cm$^{-1}$ for the pristine SiC, for the 2D-peak the line widths are 24\,cm$^{-1}$ and 31\,cm$^{-1}$, respectively. The spectrum of S4 is shifted along the intensity axis.}
    \label{Raman}
\end{figure}

Raman measurements were performed to investigate the influence of implantation induced damage on the quality of the subsequently grown graphene. Figure \ref{Raman} compares the Raman signal of sample S4 with the signal of QFMLG grown on our standard SiC substrate without implantation. The spectra show three prominent peaks at about 1340 cm$^{-1}$, 1584 cm$^{-1}$ and 2670 cm$^{-1}$ called D-, G- and 2D-band, respectively. In both spectra, the narrow 2D-line is very well described by a single Lorentzian which is typical of monolayer graphene \cite{Fer06}. Furthermore, the intensity ratio $I(D)/I(G)$, which is related to defects in graphene \cite{Tui70}, is similar. Thus, no negative influence of the implantation on the graphene quality is observed. This is complemented by the fact that the electrical properties of samples with and without implanted gate are comparable (typical hole densities are in the range of (2--6)$\cdot 10^{12}$\,cm$^{-2}$ and the typical mobility is $\mu\approx3000$\,cm$^2$/Vs).

\subsection{Exploring the limits of implantation doses}\label{exploring}
First we address the question, which minimal implantation dose is needed to fabricate a working bottom gate. As the gate layer is implanted through the insulating layer, using a $D_\mathrm{I}$ as low as possible is favorable as it minimizes implantation induced damage and therefore leakage current. Obviously for a conducting gate electrode $[D]_{\mathrm{gate}}$ has to exceed the vanadium concentration. In addition, $[D]_{\mathrm{gate}}$ has to be high enough that the gate is not depleted at high gate voltages. The depletion occurs at a $V_\mathrm{g}$, for which the induced charge carrier density $n_{\mathrm{ind}}$ becomes as large as the uncompensated donor areal density $D_{\mathrm{net}}$ in the gate layer:
\begin{equation}
n_{\mathrm{ind}}=C_{\mathrm{IPC}}V_\mathrm{g}/e\doteq D_{\mathrm{net}}\mbox{.}
\label{eq:Nuc}
\end{equation}
$C_{\mathrm{IPC}}=\varepsilon_\mathrm{SiC}\varepsilon_0/d$ is the capacitance of the bottom gate in the IPC regime with $d$ determined by TRIM simulation. For the dielectric constant of 6H-silicon carbide we used $\varepsilon_\mathrm{SiC}=9.7$ \cite{Pat70}. $D_{\mathrm{net}}$ is given by
\begin{equation}
D_{\mathrm{net}}=\int_z^{z^\prime}c\cdot[D]-[V]\mathrm{d}\zeta\mbox{,}
\label{eq:Nuc2}
\end{equation}
where $z$ or $z^\prime$ denotes the depth at which [D] exceeds or drops below $[V]$, respectively. $c$ denotes the percentage of the implanted donors which are electrically activated during the annealing step.

\begin{figure}
    \centering
        \includegraphics[width=9cm]{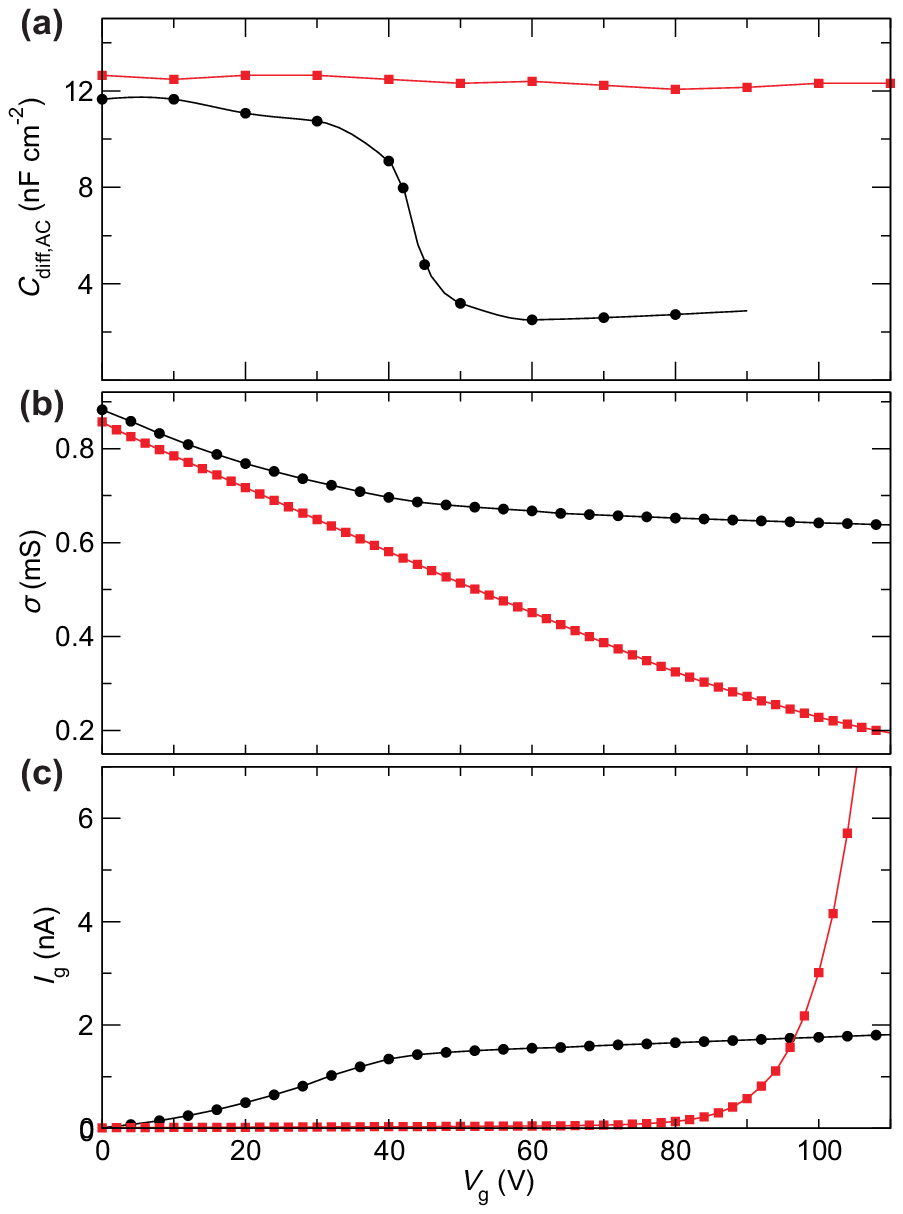}
    \caption{Gate response of S1 (circles) and S2 (squares). (a) The differential capacitance $C_{\mathrm{diff,AC}}$ of S1 shows a steep drop at $V_\mathrm{g}\approx45$\,V due to the depletion of the gate layer. In contrast, $C_{\mathrm{diff,AC}}$ of S2 is independent of gate voltage. (b) For S1 a saturation of $\sigma$ is observed for $V_\mathrm{g}>60$\,V, whereas $\sigma$ of S2 shows a linear behaviour over the whole range of $V_\mathrm{g}$. (c) Leakage current $I_\mathrm{g}$ of S1 shows a kink at $V_\mathrm{g}\approx45$\,V. $I_\mathrm{g}$ of S2 shows the typical behaviour of an insulator.}
    \label{GI5+R1}
\end{figure}

To illustrate the consequences of a depleted gate layer we compare the gate response of the differential capacitance $C_{\mathrm{diff,AC}}$, the conductivity $\sigma$ and the leakage current $I_\mathrm{g}$ of samples S1 and S2. For S1, $D_\mathrm{I}$ was chosen such that $[D]_{\mathrm{gate}}\approx 6\cdot 10^{17}\,\mathrm{cm^{-3}}\approx 6\cdot[V]$ (figure \ref{Schema}a). A roughly four times larger $D_\mathrm{I}$ was used for sample S2 (table 1). The implantation dose $D_\mathrm{I}$ is low enough that both samples operate in the IPC regime up to room temperature. The gate layer of S1 becomes depleted at $V_\mathrm{g}\approx45$\,V, which leads to a steep drop of $C_{\mathrm{diff,AC}}$ (figure \ref{GI5+R1}a). In addition, the dependence of $\sigma$ on $V_\mathrm{g}$ changes from linear to sub-linear and becomes nearly independent from $V_\mathrm{g}$ for $V_\mathrm{g}>60$\,V. Also at $V_\mathrm{g}\approx45$\,V a kink in $I_\mathrm{g}$  is observed (figure \ref{GI5+R1}b,c). This results from the drastically increased resistivity of the gate layer caused by the depletion. In this regime, if $V_\mathrm{g}$ is increased, the additional voltage $\Delta V_\mathrm{g}$ drops mainly laterally in the highly resistive gate layer. As a consequence, first the induced charge per volt in the graphene layer and thus $C_{\mathrm{diff,AC}}$ is decreased explaining the behaviour of $\sigma$. Second, the series resistance for the leakage current is increased, which shows up in the decreased slope of $I_\mathrm{g}(V_\mathrm{g})$. 
The gate voltage $V_\mathrm{g}=45\,\mathrm{V}$ at which the gate layer becomes depleted can be calculated in agreement with the experiment using equations \ref{eq:Nuc} and \ref{eq:Nuc2} with $c=0.7$, which is close to values from literature \cite{Kim95}.
 
In contrast, for sample S2 $C_{\mathrm{diff,AC}}$ is independent of $V_\mathrm{g}$ and thus no depletion of the gate layer is observed throughout the investigated gate voltage range. Therefore, $\sigma(V_\mathrm{g})$ shows the linear behaviour typical for graphene \cite{Nov05}:
\begin{equation}
\sigma(V_\mathrm{g})=e\mu n(V_\mathrm{g})=e\mu(n_0-\frac{C_{\mathrm{IPC}}}{e}V_\mathrm{g})\mbox{.}
\end{equation}
Here it is assumed that the mobility $\mu$ is independent from charge carrier density $n$, and $n_0$ is $n(V_\mathrm{g}=0)$. Note that the above formula is only valid if the Fermi level is sufficiently far away from the Dirac point so that only one type of charge carriers contributes to the current. Also $I_\mathrm{g}$ shows the expected behavior typical for an insulator with a breakdown voltage $V_\mathrm{bd}\approx 100$\,V. Note that all measurements shown were carried out with $V_\mathrm{g}<V_\mathrm{bd}$. Therefore $I_\mathrm{g}$ does not affect the measurements, as it is always orders of magnitude smaller than the source-drain current $I_\mathrm{SD}$, which is typically in the $\upmu$A range.
For S2 equations \ref{eq:Nuc} and \ref{eq:Nuc2} yield depletion of the gate layer at $V_\mathrm{g}=260$\,V, which is much larger than the breakdown voltage and is thus not reached.

The fact that for $V_\mathrm{g}<100$\,V S1 shows higher leakage currents than S2 is not attributed to the difference in implantation dose but to the local wafer quality. Probably the wafer area, of which S1 was fabricated, had a higher defect density than S2. Indeed, for $V_\mathrm{g}<V_\mathrm{bd}$ we observe similar sample to sample fluctuations of $I_\mathrm{g}$ between samples fabricated on the same chip. In contrast, the breakdown voltage $V_\mathrm{bd}$ depends on both implantation dose $D_\mathrm{I}$ and wafer quality. Larger $D_\mathrm{I}$ leads to smaller $V_\mathrm{bd}$ (e.g. for S4: $V_\mathrm{bd}\approx20$\,V). Hence, an optimal gate performance in the IPC regime is achieved when the gate voltage at which the gate layer becomes depleted and the breakdown voltage are similar. This can be controlled via $D_\mathrm{I}$.

It is remarkable that for all N-implanted samples the measured capacitance in the IPC regime $C_{\mathrm{IPC,m}}$ is in good agreement with the calculated $C_{\mathrm{IPC}}$ using a simple plate capacitor model. This shows that for samples S1 to S4 defects in the insulating layer or traps at the graphene/SiC interface have no influence on the capacitance in the IPC regime. Only for S5 using P for the gate layer the expected and the measured capacitance differ significantly (table 1). We assign this to a partial shielding of the gate effect by implantation induced defects in the insulating layer. This effect is suspected to be more important for P- than for N-implantation as the P$^+$ ions are much heavier and thus create more damage.

\begin{figure}
    \centering
        \includegraphics[width=9cm]{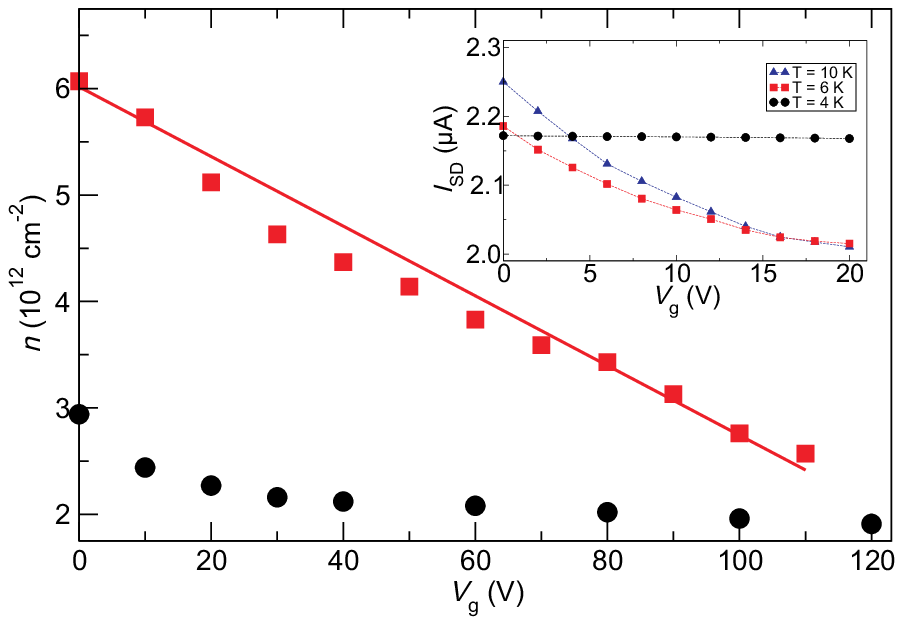}
    \caption{Change of carrier concentration $n$ upon gate voltage $V_\mathrm{g}$ for S3 (circles) and S5 (squares) at low temperatures (for S3, $T=6.0$\,K for S5 $T=4.2$\,K) and therefore in the IPC regime. For S3 only a small variation of $n$ is achieved, whereas for S5 $n$ could be changed by more than $3\cdot10^{12}$\,cm$^{-2}$. The solid line shows a linear fit to the carrier concentration of S5. The inset shows the response of the source-drain current $I_\mathrm{SD}$ on gate voltage $V_\mathrm{g}$ of S3 for different temperatures. For $T\geq6$\,K $I_\mathrm{SD}$ can be changed using the gate. For $T=4$\,K the conductivity of the gate layer freezes out and no gate response of $I_\mathrm{SD}$ is observed.}
    \label{TT}
\end{figure}

Now, we address the low-temperature limit of the bottom gate operation. Below a characteristic threshold temperature $T_\mathrm{th}$ the above used gate freezes out and becomes insulating ($T_{th}\approx75$\,K for sample S2)\cite{Wal11}. Therefore, a higher doping concentration in the gate layer is needed to maintain a finite conductivity at low $T$, ideally a quasi-metallic conductivity governed by an impurity band is used.
However, the N doping in SiC is limited to $[N]\approx 3\cdot10^{19}$\,cm$^{-3}$ \cite{Lau02}, which leads to a minimal operation temperature of $T_\mathrm{th}\approx 6\,\mathrm{K}$ (figure \ref{TT} inset).
As the maximal solubility of P in SiC is roughly one order of magnitude higher as for N \cite{Lau02} the change of the donor species allows for a higher maximal dopant concentration (for S5: $[P]_{max}=4\cdot10^{19}$\,cm$^{-3}$) and therefore a lower $T_\mathrm{th}$. We observe that the gate of S5 works at least down to $T=1.4$\,K.  
With this set of implantation parameters used for S5 a bottom gated device can be continuously used from low temperatures up to room temperature. From our perspective a further increase of $D_\mathrm{I}$ is not advantageous as high doping of the gate layer brings along several limitations. First, a high $D_\mathrm{I}$ decreases the breakdown voltage $V_\mathrm{bd}$. At elevated temperatures this is fortunately counterbalanced by the appearance of the SC regime (see below). Second, a partial shielding of the gate effect by implantation induced defects in the insulating layer is observed. For N-implanted samples this manifests as a saturation of the gate effect at higher $V_\mathrm{g}$. For P-implanted samples a partial shielding independent of $V_\mathrm{g}$ is observed. Nevertheless, P-implanted samples allow for a much wider range of accessible charge carrier densities in graphene at low $T$ than N-implanted samples (figure \ref{TT}). The capacitance $C_{\mathrm{IPC,m}}$ of sample S5 determined from the fit in figure \ref{TT} is still $C_{\mathrm{IPC,m}}=5.3$\,nF/cm$^{-2}$ at $T=4.2\,\mathrm{K}$, compared to $C_{\mathrm{IPC,m}}=9.1$\,nF/cm$^{-2}$ at $T=225\,\mathrm{K}$, where the gate electrode is well conducting but the sample still in the IPC regime.

\subsection{Modeling the different gating regimes}\label{modeling}
\begin{figure}
    \centering
        \includegraphics[width=9cm]{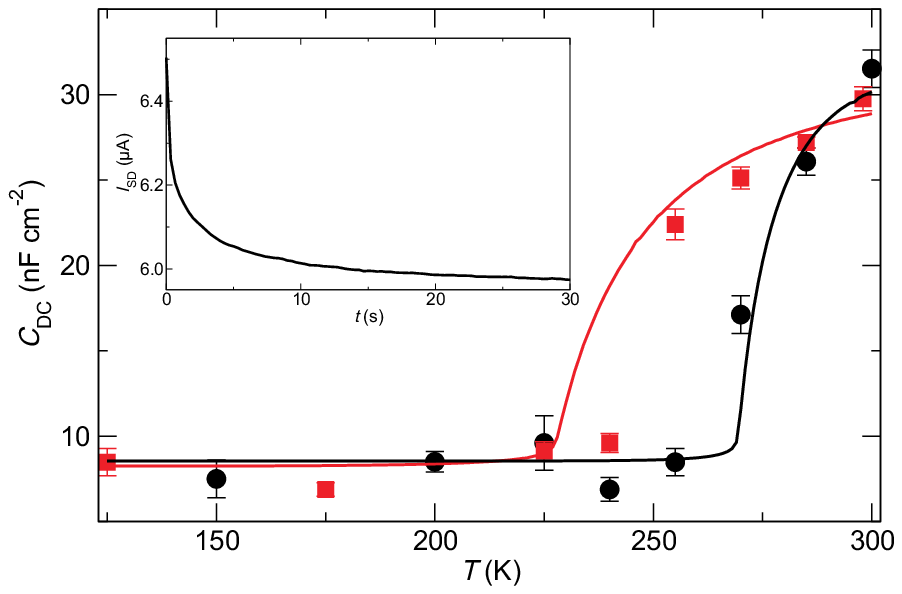}
    \caption{Temperature dependence of the capacitance $C_{\mathrm{DC}}$ of S4 (circles) and S5 (squares). Due to the transition from the IPC to the SC regime a strong increase of $C_{\mathrm{DC}}$ is observed. Solid lines show the calculated temperature dependence using the parameters from table 2. The inset shows the time-dependent response of the source-drain current $I_\mathrm{SD}$ of S5 on a change of the gate voltage from $V_\mathrm{g}=0$\,V to $V_\mathrm{g}=3$\,V at room temperature.}
    \label{C(T)}
\end{figure}

When increasing the temperature to 300\,K for sample S3 to S5 a transition from the IPC to the SC regime is observed. It manifests itself in a strong and steep increase of $C_{\mathrm{DC}}$ at the transition temperature $T_\mathrm{t}$ (figure \ref{C(T)}). We used a DC measurement to determine the capacitance $C_{\mathrm{DC}}$ in the SC regime, as the time constant of the gate response is of the order of seconds (figure \ref{C(T)} inset) and thus an AC measurement would underestimate the capacitance. For S4 we find $T_\mathrm{t}\approx 270$\,K for S5 $T_\mathrm{t}\approx 250$\,K.

 The two regimes as well as the transition can be understood by calculating the temperature dependent voltage drop of $V_\mathrm{g}$  along the insulating region between the gate electrode and the surface. In contrast to standard MOSFETs where the voltage drops homogeneously along the dielectric layer, the voltage drop in our devices is non-trivial. Due to the Schottky contact at the graphene/SiC interface $V_\mathrm{g}$ divides into two parts:
\begin{equation}
V_\mathrm{g}=V_\mathrm{SC}+V_\mathrm{IPC}\mbox{.}
\label{eq:Vg}
\end{equation}
$V_\mathrm{SC}$ drops at the Schottky contact and $V_\mathrm{IPC}$ is the standard voltage drop across a plate capacitor (figure \ref{Schema}). In the SC regime($T\gg T_\mathrm{t}$) when the resistivity of the SiC is low the whole gate voltage drops along the space charge region (SCR) of the Schottky contact ($V_\mathrm{g}=V_\mathrm{SC}$). In the IPC regime ($T\ll T_\mathrm{t}$), however, the resistivity of the insulating layer is  high and the voltage drop is equivalent to a plate capacitor ($V_\mathrm{g}=V_\mathrm{IPC}$). Here, the Schottky contact can be neglected. 
The transition from the IPC to the SC regime is caused by the different temperature dependences of $V_\mathrm{SC}(T)$ and $V_\mathrm{IPC}(T)$. To model these temperature dependences we calculate the current density across a Schottky contact $I_\mathrm{SC}$ under reverse bias, given by 
\begin{equation}
I_\mathrm{SC}(T)=A^*_\mathrm{fit} T^2\,\mbox{exp}\left(\frac{-\Phi_{\mathrm{b,eff}}}{k_\mathrm{B}T}\right)\left(\exp\left(\frac{qV_{SC}}{k_BT}\right)-1\right)\mbox{,}
\label{eq:Isc}
\end{equation}
with $\Phi_\mathrm{b,eff}$ being the effective Schottky barrier height. $A^*_\mathrm{fit}$ is the fitted effective Richardson constant of the Schottky contact. We can determine $\Phi_\mathrm{b,eff}$ and $A^*_\mathrm{fit}$ by fitting equation \ref{eq:Isc} to the measured $I_\mathrm{g}(T)$ in the SC regime. 
The resistive voltage drop $V_\mathrm{IPC}$ along the insulating layer is then obtained by
\begin{equation} 
V_\mathrm{IPC}(T)=d\rho_\mathrm{SiC}(T)I_\mathrm{SC}(T)\mbox{.}
\label{eq:VIPC}
\end{equation}
The resistivity of the insulating layer $\rho_\mathrm{SiC}(T)$ is determined with $\rho_\mathrm{SiC}(T)=\rho_0\exp(E_\mathrm{V}/(k_\mathrm{B}T))$ with $E_\mathrm{V}=0.7$\,eV being the activation energy of the vanadium acceptor level. $\rho_0$ is obtained from measuring $I_\mathrm{g}$ in forward direction of the Schottky contact at room temperature, where $I_\mathrm{g}$ is only determined by the resistivity of the insulating layer. Combining equations \ref{eq:Vg} to \ref{eq:VIPC} $V_\mathrm{SC}(T)$ can be calculated by solving
\begin{equation}
V_\mathrm{g} = V_\mathrm{SC}+d\rho_0\,\exp\left(\frac{E_\mathrm{V}}{k_\mathrm{B}T}\right)I_\mathrm{SC}(T)
\end{equation}
numerically for a given $V_\mathrm{g}$ and $T$.

The overall capacitance $C_{\mathrm{DC}}$ has two contributions. The first arises from the charge accumulated in the gate electrode, the second from the additional charge in the SCR of the Schottky contact
\begin{equation}
C_{\mathrm{DC}}=\frac{e}{V_\mathrm{g}}\left(\frac{\varepsilon_\mathrm{SiC}\varepsilon_0}{ed}V_\mathrm{IPC}+(w-w_\mathrm{min})N_\mathrm{d}\right)\mbox{,}
\end{equation}
with the thickness of the SCR $w=\sqrt{2\varepsilon_\mathrm{SiC}\varepsilon_0(V_\mathrm{SC}+V_\mathrm{bi})/(eN_\mathrm{d})}$ and $w_\mathrm{min}=w(V_\mathrm{SC}=0)$. $V_\mathrm{bi}$ is the built-in voltage of the graphene/SiC Schottky contact that was measured using X-ray photo electron spectroscopy (XPS) as described in reference \cite{Sey06}. $N_\mathrm{d}$ is the charge density in the SCR. All parameters of the model except $N_\mathrm{d}$ were determined from separate experiments (see table 2).  $N_\mathrm{d}$ was obtained by fitting the model to the measured $C_{\mathrm{DC}}(T)$ (figure \ref{C(T)}). 
\footnote{Estimating the course of the quasi-Fermi level $E_{F,n}$ for electrons in the SCR following the way presented in \cite{Res07} we find that in agreement with the fitted $N_\mathrm{d}$ the V donor level, that lies 1.5\,eV below the conduction band, is not crossing $E_{F,n}$. A crossing would lead to an increased $N_\mathrm{d}$ with $N_\mathrm{d}>[V]$ as the V would also contribute to the charge in the SCR.} 
Extracting $I_0$ and $\Phi_\mathrm{b,eff}$ simultaneously from a fit to $I_\mathrm{g}$ using equation \ref{eq:Isc} is problematic as $I_0$ is very sensitive to a change of $\Phi_\mathrm{b,eff}$. This leads to a large error in $I_0$ (table 2). To obtain $I_0$ for the model, we first chose a $\Phi_\mathrm{b,eff}$ and then determined $I_0$ using equation \ref{eq:Isc} with fixed $\Phi_\mathrm{b,eff}$. To optimize the match of the model to the data of sample S4 we varied some of the parameters within the experimental error. For the phosphorus doped sample S5 we included partial shielding of the gate effect by implantation related defects (otherwise the capacitance in the IPC regime would be too large. See table 1). The measured $\rho_0$ turned out to yield a too small transition temperature $T_\mathrm{t}$ for sample S5. We therefore used a somewhat higher value. 
The parameters used for the fits in figure \ref{C(T)} are shown in table 2. Note that $T_\mathrm{t}$ is independent of the choice of $N_\mathrm{d}$ and hence depends only on separately measured quantities.
\begin{table}
\caption{\label{tabtwo} Parameters of the model. The experimentally obtained values and the values used for the fit in figure \ref{C(T)} are shown.} 

\lineup
\begin{tabular}{@{}lccccc}
\br                              
&$V_{bi}$ (eV)&$\Phi_\mathrm{b,eff}$ (eV)&$A^*_\mathrm{fit}$ (nA K$^{-2}$)&$\rho_0$ ($\Omega$m)&$N_\mathrm{d}$ ($10^{16}$\,cm$^{-3}$)\cr 
\mr
S4 meas.&$0.9\pm0.1$&$0.13\pm0.03$&$(1.0\pm1.3)\cdot 10^{-3}$&$(2.2\pm1.9)\cdot 10^{-4}$&-\cr
S4 fit&$0.9$&$0.14$&$1.1\cdot 10^{-3}$&$9.5\cdot 10^{-5}$&3.1\cr 
S5 meas.&$0.9\pm0.1$&$0.56\pm0.03$&$(1.2\pm1.4)\cdot 10^{4}$&$(1.2\pm0.8)\cdot 10^{-4}$&-\cr 
S5 fit&$0.9$&$0.53$&$4.3\cdot 10^{3}$&$3.5\cdot 10^{-4}$&4.5\cr 

\br
\end{tabular}
\end{table}
\newline
In the SC regime the differential capacitance of the gate depends on $V_\mathrm{g}$. With increasing $V_\mathrm{g}$ the SCR becomes larger and thus the differential capacitance decreases. Hence, the induced charge carrier density $n_{\mathrm{ind}}$ becomes sub-linear. Figure \ref{n_ind}a shows that for small $V_\mathrm{g}$ we find excellent agreement with the calculated $n_{\mathrm{ind}}$ at room temperature using the textbook formula
\begin{equation}
n_{\mathrm{ind}}(V_\mathrm{g})=\sqrt{2\varepsilon_\mathrm{SiC}\varepsilon_0N_\mathrm{d}/e}\left(\sqrt{V_\mathrm{g}+V_\mathrm{bi}}-\sqrt{V_\mathrm{bi}}\right)\mbox{.}
\label{eq:nind}
\end{equation}
For $V_\mathrm{g}>9$\,V the measured $n_{\mathrm{ind}}$ is slightly smaller as the theoretically predicted values. Probably this is caused by the filling of traps at high electric field that was already observed in \cite{Wal11}.\newline

\subsection{Controlling the transition to the SC regime with light}\label{controlling}
\begin{figure}
    \centering
    		\includegraphics[width=9cm]{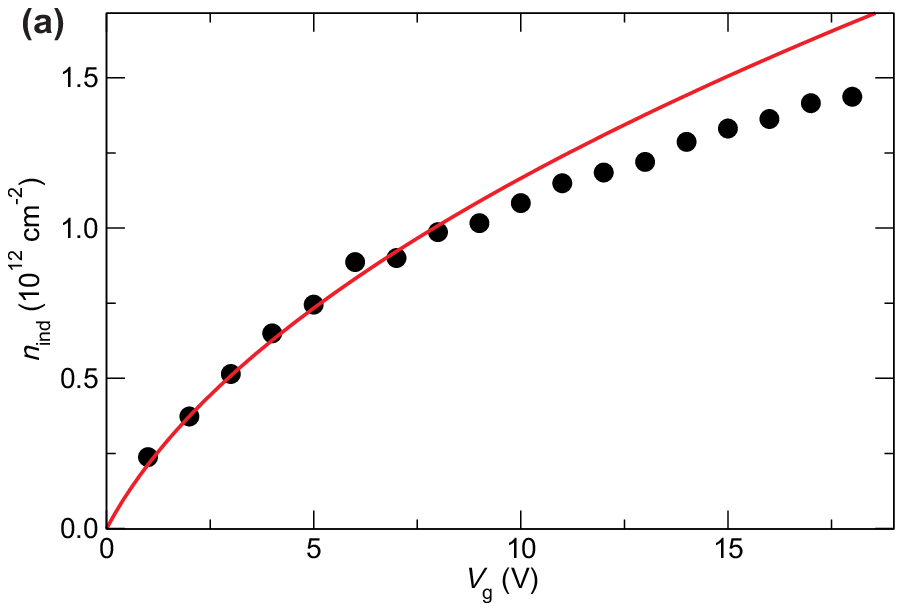}
    		\vspace{1cm}
        \includegraphics[width=9cm]{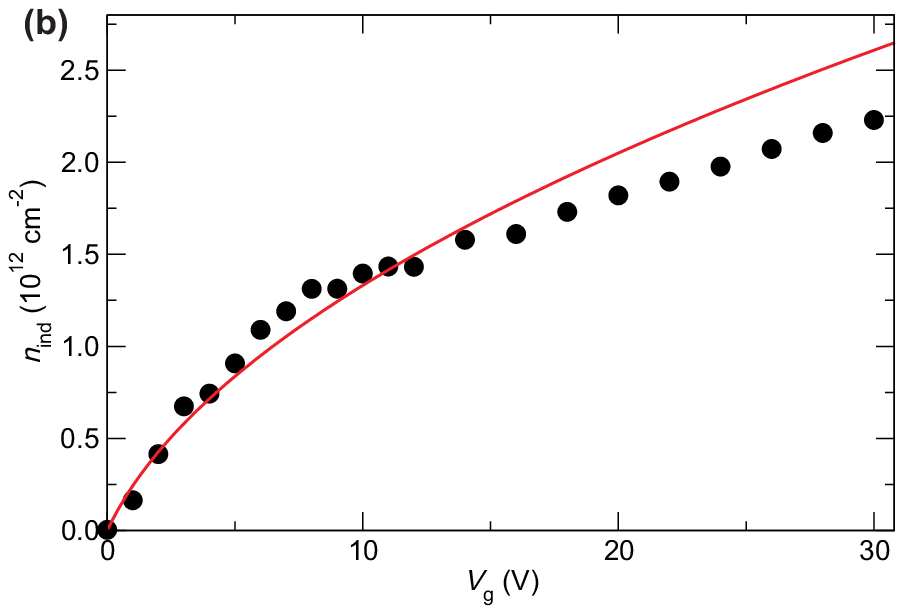}
    \caption{Measured (circles) and calculated (solid line) dependence of the induced charge carrier density $n_{\mathrm{ind}}$ on gate voltage for S4. (a) In  the SC regime at room temperature. (b) With UV light illumination at $T=35$\,K. The UV light allows to extend the SC regime to low temperatures.}
    \label{n_ind}
\end{figure}

To extend the SC regime to low temperatures we apply ultra violet (UV) light with a photon energy larger than the SiC band gap ($E_\mathrm{g}=3.0$\,eV). The UV illumination creates additional free charge carriers in the insulating layer and its resistivity is decreased similar to the already discussed case of elevated temperatures. Figure \ref{n_ind}b shows that with UV illumination $n_{\mathrm{ind}}(V_\mathrm{g})$ has the sub-linear behaviour typical for the SC regime also at $T=35$\,K. Again for small $V_\mathrm{g}$ the measured $n_{\mathrm{ind}}(V_\mathrm{g})$ is very well described by equation \ref{eq:nind}. From the fits shown in figure \ref{n_ind}a and \ref{n_ind}b we obtain $N_\mathrm{d}=2.3\cdot10^{16}$\,cm$^{-3}$ and $N_\mathrm{d}=2.9\cdot10^{16}$\,cm$^{-3}$, respectively. This is slightly smaller than $N_\mathrm{d}$ obtained from the fit in figure \ref{C(T)}. We assign this to a slow degradation of the gate coupling as applying high gate voltages probably creates additional defects that shield the gate effect.

\begin{figure}
    \centering
        \includegraphics[width=9cm]{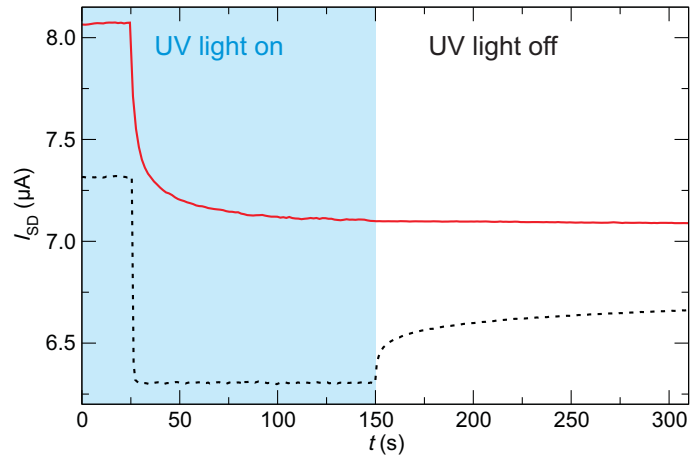}
    \caption{Gate response of the source-drain current $I_\mathrm{SD}$ of S4 at $T=100$\,K (dashed line) and $T=35$\,K (solid line). At the beginning of the measurement the sample is illuminated with UV light. At $t=25$\,s the gate voltage is changed from $V_\mathrm{g}=0$\,V to $V_\mathrm{g}=3$\,V and at $t=150$\,s the UV light is switched off. At $T=100$\,K the sample undergoes a transition back to the IPC regime whereas at $T=35$\,K it remains in the SC regime.}
    \label{UVoff}
\end{figure}

The overall picture is fully confirmed by the following observation: when switching off the UV light at low temperatures the sample remains in the SC regime. This is a consequence of the extremely high resistivity of the insulating layer at low temperatures without UV illumination. Under these conditions the charge can not rearrange to the IPC distribution.
 Figure \ref{UVoff} shows the response of $I_\mathrm{SD}$ on a change of $V_\mathrm{g}$ from $V_\mathrm{g}=0$\,V to $V_\mathrm{g}=3$\,V for $T=100$\,K and $T=35$\,K with the sample being illuminated with UV light. For both temperatures the sample is in the SC regime due to the UV light and the observed gate response of $I_\mathrm{SD}$ is similar. Upon switching the UV light off, at $T=100$\,K $I_\mathrm{SD}$ increases indicating a reduction of the gate coupling and a transition from the SC to the IPC regime. At $T=35$\,K, however, no change in $I_\mathrm{SD}$  is observed and the sample stays in the SC regime for at least several hours. It is a nice property of the system that low temperatures measurements in the favorable SC regime are possible without the disturbing influence of UV light.

\section{Conclusion}
We present a scheme to create a bottom gate in epitaxial graphene devices by ion implantation prior to graphene growth. The SiC takes over the functionality of both the conducting gate layer and the insulating dielectric. We have shown that substantial changes in charge carrier densities can be achieved. In this paper we give a detailed description of the influence of various doping profiles on the graphene/SiC heterojunction. In particular we elucidate the practical limits of implantation doses. Furthermore a nearly parameter-free model is developed which describes the underlying physics with high accuracy. We find two regimes for the gate capacitance with the Schottky regime being more efficient. Using UV light we can force the system to the SC regime even at cryogenic temperatures. The implanted bottom gate allows qualitatively new experiments using epitaxial graphene.    

\ack
We gratefully acknowledge support by the Cluster of Excellence `Engineering of Advanced Materials' (www.eam.uni-erlangen.de) at the Friedrich-Alexander-Universit\"at Erlangen-N\"urnberg.

\newpage
\bibliographystyle{unsrt}
\bibliography{graphene}

\newpage

\end{document}